\documentclass{article}
\usepackage{times}
\usepackage{amsfonts}
\usepackage{graphicx}
\begin{document}
\noindent
{\Large GENERALISED COMPLEX GEOMETRY IN THERMODYNAMICAL FLUCTUATION THEORY}
\vskip.5cm
\noindent
{\bf P. Fern\'andez de C\'ordoba}$^{a}$ and {\bf  J.M. Isidro}$^{b}$\\
Instituto Universitario de Matem\'atica Pura y Aplicada,\\ Universidad Polit\'ecnica de Valencia, Valencia 46022, Spain\\
${}^{a}${\tt pfernandez@mat.upv.es}, ${}^{b}${\tt joissan@mat.upv.es}
\vskip.5cm
\noindent
{\bf Abstract} We present a brief overview of some key concepts in the theory of generalised complex manifolds.  This new geometry interpolates, so to speak, between symplectic geometry and complex geometry. As such it provides an ideal framework to analyse thermodynamical fluctuation theory in the presence of gravitational fields. To illustrate the usefulness of generalised complex geometry, we examine a simplified version of the Unruh effect: the thermalising effect of gravitational fields on the Schroedinger wavefunction.  


\section{Introduction}\label{unno}

The theory of thermodynamical fluctuations provides a solid link between macroscopic and microscopic physics.  Classical fluctuation theory \cite{CALLEN} often sheds light on counterintuitive quantum--mechanical phenomena, thus helping to bridge the gap between the classical world and the quantum world. For example, Heisenberg's uncertainty principle can be nicely illustrated resorting to the theory of Gaussian fluctuations around thermal equilibrium \cite{VELAZQUEZ1}.

On the other hand, the theory of thermodynamical fluctuations can be recast using the geometric language of differential manifolds \cite{BRAVETTI1, BRAVETTI2, QUEVEDO, RAJEEV1, RAJEEV2, RUPP1, VELAZQUEZ2}. This reexpression of a physical discipline in more abstract mathematical language goes a long way beyond a mere rewriting of the concepts involved. It renders the theory more versatile, enlarging its scope. Moreover, since the advent of Einstein's general relativity a century ago, (pseudo) Riemannian geometry belongs to the technical skills that any physicist has to master (at least at a working level). This places (pseudo) Riemannian geometry  at a vantage point. In the opposite direction ({\it i.e.}\/, thermodynamics as applied to geometry) one should mention at least two developments. The first one is a whole body of knowledge on the thermodynamics of black holes \cite{BCH, RUPP2}. More recently, the reexpression of Einstein's relativity as a thermodynamics \cite{PADDY2, PADDY3} has had far--reaching consequences for our understanding of spacetime.

Here we would like to report on another recent development in geometry with implications on the thermodynamics of fluctuations: the theory of generalised complex manifolds \cite{GUALTIERI, HITCHIN}. 

In trying to understand the thorny relationship between gravity and the quantum \cite{CALMET, ELZE3, THOOFT, MATONE} it has been argued that
gravity acts dissipatively on quantum systems \cite{PENROSE}. Specifically, {\it in the presence of a gravitational field, thermal fluctuations become indistinguishable from quantum fluctuations}\/ \cite{PADDY1, SMOLIN1, SMOLIN2}. This raises the fundamental question: How is one to treat thermal {\it and}\/ quantum fluctuations on the same footing? Is it altogether possible? We will see here that generalised complex manifolds provide one viable answer to this question, one that appears not to have been explored yet in the geometrical approach to thermodynamics.

\section{Geometry and fluctuations}\label{ddos}

\subsection{Riemannian geometry}\label{riemomo}

As a very elementary example, consider a thermodynamical system in an equilibrium state described by the following variables: temperature $T$,  pressure $P$ and  volume $V$.  In the Gaussian approximation, choosing $T$ and $V$ as independent variables, the probability $W$ of a fluctuation $\Delta T$, $\Delta V$ around equilibrium is given by \cite{LANDAU}
\begin{equation}
W=W_0\exp\left[-\frac{C_V}{2k_BT^2}\Delta T^2+\frac{1}{2k_BT}\left(\frac{\partial P}{\partial V}\right)_T\Delta V^2\right].
\label{nned}
\end{equation}
The thermodynamic inequalities $C_V>0$ and $(\partial P/\partial V)_T<0$ ensure that the argument of the above exponential is negative definite. This suggests considering the following (positive definite) Riemannian metric on the 2--dimensional manifold coordinatised by $T,V$:
\begin{equation} 
{\rm d}s^2:=\frac{C_V}{2k_BT^2}{\rm d} T^2-\frac{1}{2k_BT}\left(\frac{\partial P}{\partial V}\right)_T{\rm d}V^2=:g_{ij}{\rm d}x^{i}{\rm d}x^j.
\label{metrik}
\end{equation}
The metric coefficients $g_{ij}$ are of course $(T,V)$--dependent functions. This Riemannian structure encodes all the relevant information. For example, 
the average value $\langle f(T,V)\rangle$ of an arbitrary function $f=f(T,V)$,
\begin{equation}
\langle f(T,V)\rangle = Z^{-1}\int f(T,V)\exp\left(-g_{TT}T^2-g_{VV}V^2\right)\sqrt{g}\,{\rm d}T{\rm d}V,
\label{quinto}
\end{equation}
where $Z:=\int\sqrt{g}\,\exp\left(-g_{TT}T^2-g_{VV}V^2\right){\rm d}T{\rm d}V$, naturally involves the metric. The role of Riemannian geometry in fluctuation theory is well known and has been reviewed at length in ref. \cite{RUPP1}.

\subsection{Symplectic geometry}\label{gemo}

As our starting point here we will consider a certain thermodynamical system in equilibrium, in order to arrive at a corresponding symplectic structure. 

Again in the Gaussian approximation, the probability $W$ of a fluctuation $\Delta P$, $\Delta V$, $\Delta T$, $\Delta S$ is given by \cite{LANDAU}
\begin{equation}
W=W_0\exp\left[-\frac{1}{2k_BT}(-\Delta P\Delta V+\Delta T\Delta S)\right].
\label{occhiali}
\end{equation}
Assume an equation of state $F(P,V,T)=0$ that can be solved for the temperature to obtain $T=g(P,V)$. For simplicity let us consider an ideal gas, $PV=S_0T$:
\begin{equation}
W=W_0\exp\left[-\frac{1}{2k_B}\left(-S_0\frac{\Delta P\Delta V}{PV}+\frac{\Delta T\Delta S}{T}\right)\right].
\label{zuhtto}
\end{equation}
It is convenient to define the dimensionless variables
\begin{equation}
p_1:=-\ln\left(\frac{P}{P_0}\right),\quad q_1:=\ln\left(\frac{V}{V_0}\right),\quad p_2:=\ln\left(\frac{T}{T_0}\right), \quad q_2:=\frac{S}{S_0},
\label{caloret}
\end{equation}
where $P_0$, $V_0$ and $T_0$ are reference values. Then Eq. (\ref{zuhtto}) becomes
\begin{equation}
W=W_0\exp\left[-\frac{S_0}{2k_B}\left(\Delta p_1\Delta q_1+\Delta p_2\Delta q_2\right)\right].
\label{fertig}
\end{equation}
We can regard $q_1$ and $q_2$ as coordinates on a thermodynamical configuration space $\mathbb{S}$, with $p_1$ and $p_2$ as their conjugate momenta. Thus the $q_1$, $p_1$, $q_2$, $p_2$ are Darboux coordinates for the symplectic form
\begin{equation}
\omega={\rm d}p_1\wedge{\rm d}q_1+{\rm d}p_2\wedge{\rm d}q_2.
\label{pohhsdg}
\end{equation}
In this way we identify $\Delta p_1\Delta q_1+\Delta p_2\Delta q_2$ in Eq. (\ref{fertig}) as the symplectic area of a 2--dimensional surface $\mathbb{F}$ induced by the fluctuation:
\begin{equation}
\Delta p_1\Delta q_1+\Delta p_2\Delta q_2=\int_{\mathbb{F}}\left({\rm d}p_1\wedge{\rm d}q_1+{\rm d}p_2\wedge{\rm d}q_2\right).
\label{drrkk}
\end{equation}
{}Finally substituting Eq. (\ref{drrkk}) into Eq. (\ref{fertig}) we find
\begin{equation}
W=W_0\exp\left(-\frac{S_0}{2k_B}\int_{\mathbb{F}}\omega\right),
\label{lla}
\end{equation}
{\it i.e.}\/, the probability of this thermal fluctuation is proportional to the exponential of the symplectic area of the fluctuation surface $\mathbb{F}$.

The importance of symplectic structures in {\it classical}\/ mechanics is widely recognised and need hardly be recalled \cite{ARNOLD}. In fact not just Riemannian geometry, but also symplectic geometry, pertains to the realm of {\it thermal}\/ fluctuations: the first law of thermodynamics endows the thermodynamic phase space with a contact structure, which includes symplectic geometry as a sub--case \cite{BRAVETTI1, BRAVETTI2, RAJEEV1, RAJEEV2}.

A real $2n$--dimensional manifold $\mathbb{M}$ is symplectic if there exists a closed, nondegenerate, rank 2 antisymmetric tensor field $\omega_{ij}$ defined everywhere on $\mathbb{M}$. Let $x^{i}$ be local coordinates around $x\in\mathbb{M}$, so $\omega=\frac{1}{2}\omega_{ij}{\rm d}x^{i}\wedge{\rm d}x^j$ with $\omega_{ji}=-\omega_{ij}$. Since the matrix $\omega_{ij}$ is nonsingular, an inverse $\pi^{jk}$ exists such that $\omega_{ij}\pi^{jk}=\delta_{i}^{k}$. The Poisson brackets of two functions $f,g$ are defined  as $\{f,g\}:=\pi^{jk}\partial_jf\partial_kg$, and the integrability condition ${\rm d}\omega=0$ turns out to be equivalent to the Jacobi identity for these Poisson brackets.

In this way the following symplectic analogue of Eq. (\ref{quinto}) allows one to compute the average value $\langle f \rangle$ of the function $f$ on $\mathbb{M}$:
\begin{equation}
\langle f\rangle= Z^{-1}\int_{\mathbb{M}}f\exp\left(-\omega\right).
\label{septimo}
\end{equation}
Above, the exponential ${\rm e}^{-\omega}$ is defined by Taylor expansion, powers being taken with respect to the wedge product. Then the $2n$--dimensionality of the symplectic manifold picks out just one differential form that can be integrated against $\mathbb{M}$, namely the $2n$--form $(-1)^n\omega^n/n!$; all other terms in the Taylor expansion give a vanishing contribution when integrated. The factor $(-1)^n/n!$ has been included in the normalisation $Z$. As had to be the case, this average involves the data concerning the symplectic structure on $\mathbb{M}$.

One can also regard a symplectic structure as providing an isomorphism from the tangent space $T_x\mathbb{M}$ into the cotangent space $T_x^*\mathbb{M}$ at each $x\in\mathbb{M}$. Specifically, the tangent vector $X=X^{i}\partial_i$ is mapped into the 1--form $\omega(X)=\xi=\xi_{i}{\rm d}x^{i}$, with  $\xi_i=\omega_{ij}X^{j}$. This viewpoint motivates the following definition (equivalent to the above, but more useful for later applications): a symplectic structure over a $2n$--dimensional manifold $\mathbb{M}$ is an isomorphism $\omega_x$ between the tangent and the cotangent fibres over each point $x\in\mathbb{M}$,
\begin{equation} 
\omega_x\colon T_x\mathbb{M}\longrightarrow T^*_x\mathbb{M},
\label{kksympww}
\end{equation}
such that, under the operation of taking the linear dual (denoted by an asterisk),
\begin{equation}
\omega_x^*=-\omega_x,\qquad \forall x\in\mathbb{M}.
\label{nanu}
\end{equation}
Moreover, the integrability condition ${\rm d}\omega=0$ must be satisfied.

\subsection{Complex geometry and K\"ahler geometry}\label{caler}

Informally one could say that the imaginary unit is the hallmark of {\it quantum}\/ mechanics. That ${\rm i}=\sqrt{-1}$ pertains to the quantum world has been very interestingly argued recently in refs. \cite{KAUFFMAN1, KAUFFMAN2}. More standard arguments have been known for long; such are 
the heat equation in imaginary time ${\rm i}t$, or the fact that quantum commutators $[\cdot\,,\cdot]$ formally equal $\sqrt{-1}$ times classical Poisson brackets $\{\cdot\,,\cdot\}$. Here we will briefly recall the role played by complex structures in the theory of coherent states \cite{LISBON, PERELOMOV}. 
 
Let $\mathbb{M}$ be a real $2n$--dimensional phase space endowed with the symplectic form $\omega$. For simplicity let us also assume that $\mathbb{M}$ admits a holomorphic atlas compatible with the symplectic structure (this compatibility condition is called  {\it the K\"ahler property}\/). In plain words, the real and imaginary parts of the holomorphic coordinates $z^j$ are Darboux coordinates for $\omega$ (here assumed dimensionless for simplicity):
\begin{equation}
z^j=\frac{1}{\sqrt{2}}\left(q^j + {\rm i} p_j\right),  \qquad j=1,\ldots, n.
\label{koomfgerl}
\end{equation}
Upon quantisation, the Darboux coordinates $q^j$ and $p_j$ become operators $Q^j$ and $P_j$ on Hilbert space satisfying the Heisenberg algebra 
$[Q^j, P_k]={\rm i}\delta^{j}_{k}$. Creation and annihilation operators are defined in the standard fashion: $A_j ^{\dagger}:=(Q^j-{\rm i} P_j)/\sqrt{2}$, $A_j:=(Q^j+{\rm i} P_j)/\sqrt{2}$, and quantum excitations are measured with respect to a vacuum state $\vert 0\rangle$ satisfying
$A_j\vert 0\rangle = 0$,  for all $j=1, \ldots, n$. Coherent states $\vert z^j\rangle$ are eigenvectors of $A_j$, the eigenvalues being the holomorphic coordinates (\ref{koomfgerl}):
\begin{equation}
A_j\vert z^j\rangle=z^j\vert z^j\rangle,\qquad j=1,\ldots, n.
\label{poean}
\end{equation}
(No sum over $j$ implied). In order to illustrate our point let us consider a 1--dimensional harmonic oscillator. The expectation value of the Hamiltonian operator $H=A^\dagger A+1/2$ in the state $\vert z\rangle$ equals $\langle z\vert H\vert z\rangle=\vert z\vert^2+1/2$. Since the energy fluctuation in the state $\vert z\rangle$ equals  
\begin{equation}
(\Delta H)_z=\vert z\vert, \qquad z\in\mathbb{C},
\label{thad}
\end{equation}
the relative fluctuation goes, for large enough $\vert z\vert$, like 
\begin{equation}
\frac{(\Delta H)_z}{\langle z\vert H\vert z\rangle}\simeq\frac{1}{\vert z\vert}, \qquad \vert z\vert\to\infty.
\label{marc}
\end{equation}
But $1/\vert z\vert$ is the inverse of the square root of the K\"ahler potential $K(z,\bar z):=\vert z\vert^2$ for the Euclidean metric on the complex plane $\mathbb{C}$. This simple example illustrates the important role played by complex manifolds in the quantum theory.

Every complex manifold $\mathbb{M}$ admits a (positive definite) Hermitian metric $h_{i j}{\rm d} \bar z^{i}{\rm d}z^j$ that is compatible with the complex structure \cite{KOBAYASHI}. Then an analogue of Eqs. (\ref{quinto}) and (\ref{septimo}) gives us the average value $\langle f\rangle$ of a function $f$ on $\mathbb{M}$:
\begin{equation}
\langle f\rangle=Z^{-1}\int_{\mathbb{M}}f\exp\left(-h_{ij}\bar z^{i}z^j\right)\sqrt{h}\,\prod_{k=1}^n{\rm d}\bar z^k\wedge{\rm d}z^k.
\label{severus}
\end{equation}
The normalisation $Z$ includes all factors of ${\rm i}=\sqrt{-1}$ coming from the volume element, and $h:=\vert\det h_{i j}\vert$. As had to be the case, this average involves the data concerning the complex structure on $\mathbb{M}$.

{}Formally, a complex structure $J$ over a real $2n$--dimensional manifold $\mathbb{M}$ is an endomorphism of the tangent fibre over each point $x\in\mathbb{M}$
\begin{equation}
J_x\colon T_x\mathbb{M}\longrightarrow T_x\mathbb{M}
\label{tratra}
\end{equation}
satisfying 
\begin{equation}
J_x^2=-{\bf 1},\qquad \forall x\in\mathbb{M},
\label{ketef}
\end{equation}
as well as the integrability condition that the Nijenhuis tensor $N$ vanish identically.  (We will not write down the Nijenhuis tensor explicitly; see ref. \cite{KOBAYASHI} for details). Roughly speaking, Eq. (\ref{ketef}) expresses the existence of the imaginary unit ${\rm i}=\sqrt{-1}$ locally around the point $x\in\mathbb{M}$. The integrability condition $N=0$ ensures that the complex coordinates thus constructed locally truly transform holomorphically across different coordinate patches on the manifold $\mathbb{M}$.  (The K\"ahler property  assumed in Eq. (\ref{koomfgerl}) above is an additional hypothesis, that an arbitrary complex manifold may, but need not, satisfy in general).

\subsection{Generalised complex geometry}\label{jichin}

Our original motivation was the statement \cite{PADDY1, SMOLIN1, SMOLIN2} that, {\it in the presence of a gravitational field, quantum fluctuations become indistinguishable from thermal fluctuations}\/. We have argued that thermal fluctuations are associated with symplectic structures, while quantum fluctuations come along with complex structures. How, then, is one to treat thermal {\it and}\/ quantum fluctuations on the same footing?  This is trivially achieved by those phase spaces $\mathbb{M}$ that qualify as  K\"ahler manifolds. However, the K\"ahler condition is very restrictive: not only does $\mathbb{M}$ have to be simultaneously complex {\it and}\/ symplectic; these two independent structures also have to be compatible.

In refs. \cite{BRAVETTI1, BRAVETTI2} the geometry of the thermodynamic phase space (including fluctuations) results in a para--Sasakian manifold, which is the contact--geometry equivalent of a K\"ahler manifold in symplectic geometry. This means that if one restricts to a proper even--dimensional subspace, the geometry is indeed that of a K\"ahler manifold. This geometry achieves the goal of treating thermal and quantum fluctuations on the same footing. 

Generalised complex structures (GCS for short) also achieve the goal of providing a unified framework for thermal and quantum fluctuations. The following is a brief summary of GCS extracted from ref. \cite{GUALTIERI}, duly taylored to meet our needs. For simplicity we prefer to work locally around a point $x\in\mathbb{M}$. Global issues can be taken care of by the corresponding integrability conditions, to be mentioned along the way whenever necessary. For our purposes the $2n$--dimensional manifold $\mathbb{M}$ is assumed to be a phase space, that is, $\mathbb{M}=T^*\mathbb{S}$, for a certain $n$--dimensional configuration space $\mathbb{S}$.

Rather than considering the fibres $T_x\mathbb{M}$ or $T_x^*\mathbb{M}$ separately, in generalised complex geometry one considers their direct sum: over each point $x\in\mathbb{M}$ one erects the fibre $T_x\mathbb{M}\oplus T_x^*\mathbb{M}$. The total space of the bundle so constructed is $6n$--dimensional: $2n$ dimensions for the base $\mathbb{M}$, $4n$ dimensions for the fibre. 

An inner product is defined on the fibre $T_x\mathbb{M}\oplus T^*_x\mathbb{M}$: 
\begin{equation}
\langle X+\xi,Y+\eta\rangle:=\frac{1}{2}\left(\xi(Y)+\eta(X)\right).
\label{fghrutt}
\end{equation}
Above, $X,Y\in T_x\mathbb{M}$ are tangent vectors, while $\xi, \eta\in T_x^*\mathbb{M}$ are 1--forms, all evaluated at $x\in\mathbb{M}$. It turns out that this inner product is pseudo--Riemann with signature $(2n,2n)$. Hence the Lie group $SO(2n,2n)$ acts on $T_x\mathbb{M}\oplus T^*_x\mathbb{M}$ by isometries. It is convenient to block--decompose the Lie algebra $so(2n,2n)$ as follows:
\begin{equation}
\left(\begin{array}{cc}
A&\beta\\
B&-A^*
\end{array}\right).
\label{nenwfervf}
\end{equation}
The diagonal blocks $A$ and $A^*$ are endomorphisms of their respective (sub)fibres, $A\in{\rm End}(T_x\mathbb{M})$ and  $A^*\in{\rm End}(T^*_x\mathbb{M})$, while the offdiagonal blocks $B$ and $\beta$ connect these two (sub)fibres as per
\begin{equation}
B:T_x\mathbb{M}\longrightarrow T_x^*\mathbb{M}, \qquad \beta:T_x^*\mathbb{M}\longrightarrow T_x\mathbb{M}.
\label{barbieri}
\end{equation}
Moreover, upon taking the dual we have $B^*=-B$, $\beta^*=-\beta$. This antisymmetry allows us to regard the block $B$ as a 2--form in $\Lambda^2T_x^*\mathbb{M}$ if we set
\begin{equation}
B(X)=i_XB. 
\label{adelante}
\end{equation}
For illustrative purposes let us express Eq. (\ref{adelante}) in local coordinates $x^{i}$ around a point $x\in\mathbb{M}$, so $B$ becomes the matrix $B_{ij}$. Given the vector $X=X^j\partial_j\in T_x\mathbb{M}$, the object $i_XB$ is defined to be the covector whose components are $B_{ij}X^j\in T_x^*\mathbb{M}$. We see that this is exactly the way a symplectic form $\omega$ behaves. Since $\omega$ can be regarded as an element of $\Lambda^2T_x^*\mathbb{M}$, so can $B$. (Contrary to $\omega$, however, $B$ need neither be closed nor nondegenerate).

The particular isometries of the fibre $T_x\mathbb{M}\oplus T_x^*\mathbb{M}$ obtained by setting $A=0=\beta$ in Eq. (\ref{nenwfervf}) and exponentiating,
\begin{equation}
{\rm exp}\left(\begin{array}{cc}
0&0\\
B&0
\end{array}\right)=\left(\begin{array}{cc}
{\bf 1}&0\\
B&{\bf 1}
\end{array}\right),
\label{kkzz}
\end{equation}
are the pseudo--orthogonal transformations
\begin{equation}
X+\xi\longrightarrow X+\xi+i_XB.
\label{azaazo}
\end{equation}
The isometries (\ref{azaazo}), called {\it B--transformations}\/, will play an important role.

A {\it generalised complex structure}\/ over $\mathbb{M}$, denoted ${\cal J}$, is an endomorphism of the fibre over each $x\in \mathbb{M}$,
\begin{equation}
{\cal J}_x\colon T_x\mathbb{M}\oplus T^*_x\mathbb{M}\longrightarrow T_x\mathbb{M}\oplus T_x^*\mathbb{M},
\label{buco}
\end{equation}
such that the following two conditions hold.  First,
\begin{equation}
{\cal J}_x^2=-{\bf 1}, \qquad \forall x\in\mathbb{M}.
\label{fotini}
\end{equation}
Second, 
\begin{equation}
{\cal J}_x^*=-{\cal J}_x,\qquad \forall x\in\mathbb{M}.
\label{kaspp}
\end{equation}
The above two conditions are formulated locally around any $x\in\mathbb{M}$; as usual they need not be compatible with changes of coordinate charts on $\mathbb{M}$. The {\it Courant integrability condition}\/, whose validity we will henceforth assume without stating its contents explicitly, ensures this compatibility;  see refs. \cite{GUALTIERI, HITCHIN} for details. 

Comparing now Eqs. (\ref{kaspp}) and (\ref{nanu}), we are led to the particular case when ${\cal J}$ at $x\in \mathbb{M}$ is given by
\begin{equation}
{\cal J}_{\omega_x}=\left(\begin{array}{cc}
0&-\omega_x^{-1}\\
\omega_x&0\end{array}\right),
\label{truienm}
\end{equation}
where $\omega$ is a symplectic form. One says that this ${\cal J}_{\omega}$ defines a GCS 
{\it of symplectic type}\/. 

Similarly, the comparison of Eqs. (\ref{fotini}) and (\ref{ketef}) suggests the particular case of a GCS given by
\begin{equation}
{\cal J}_{J_x}=\left(\begin{array}{cc}
-J_x&0\\
0&J^*_x\end{array}\right),
\label{peliniii}
\end{equation}
where $J$ is a complex structure. We say that the above ${\cal J}_{J}$ defines a GCS {\it of complex type}\/. 

Furthermore, GCS succeed at interpolating between the above opposite types, the {\it symplectic type}\/ and the {\it complex type}\/; let us explain this more carefully. A point $x\in\mathbb{M}$ is said to be {\it regular}\/ if it possesses a neighbourhood ${\cal N}_x$ on which there exists a Poisson structure $\omega^{-1}$ with constant rank. In a neighbourhood ${\cal N}_x$ of any regular point $x\in\mathbb{M}$ one can define a diffeomorphism and a $B$--transformation,  the combined action of which maps ${\cal N}_x$  into the product ${\cal C}_x\times {\cal R}_{x}\subset \mathbb{C}^k\times \mathbb{R}^{2n-2k}$. Here ${\cal C}_x$ is an open set within the standard complex manifold $\mathbb{C}^k$, and ${\cal R}_{x}$ is an open set within the standard symplectic manifold $\mathbb{R}^{2n-2k}$. The nonnegative integer $k$ is called the {\it type}\/ of the GCS ${\cal J}$, the limiting cases of Eqs.  (\ref{truienm}) and (\ref{peliniii}) respectively corresponding to $k=0$ and $k=n$. As described in refs. \cite{GUALTIERI, HITCHIN}, the type $k$ need not be constant across $\mathbb{M}$: it may vary from one point to another in $\mathbb{M}$.

In plain words, any generalised complex manifold factorises {\it locally}\/ as the product of a complex manifold times a symplectic manifold. 

{}Finally assume that $\mathbb{M}$ is a linear space. Then any generalised complex structure of type $k=0$ is the $B$--transform of a symplectic structure. This means that any generalised complex structure of type $k=0$ can be written as
\begin{equation}
{\rm e}^{-B}{\cal J}_{\omega}{\rm e}^{B}=
\left(\begin{array}{cc}
-\omega^{-1}B & -\omega^{-1}\\
\omega + B\omega^{-1}B & B\omega^{-1}\end{array}\right)
\label{baspp}
\end{equation}
for a certain 2--form $B$; use has been made of Eqs. (\ref{kkzz}) and (\ref{truienm}). Similarly any generalised complex structure of type $k=n$ over a linear manifold $\mathbb{M}$ is the $B$--transform of a complex structure,
\begin{equation}
{\rm e}^{-B}{\cal J}_{J}{\rm e}^{B}=\left(\begin{array}{cc}
-J & 0\\
BJ+J^*B & J^*\end{array}\right),
\label{merde}
\end{equation}
after using Eqs. (\ref{kkzz}) and (\ref{peliniii}). When $\mathbb{M}$ is an arbitrary smooth manifold, not necessarily a linear space, statements (\ref{baspp}) and (\ref{merde}) remain basically true, with some minor modifications required; see refs. \cite{GUALTIERI, HITCHIN} for details.

The consequences of the above become immediately apparent. Let us for simplicity assume that the type $k$ is constant across $\mathbb{M}$. Then any GCS with an extremal value of $k$, {\it i.e.}\/, either $k=0$ or $k=n$, can always be reduced to the corresponding canonical form (\ref{truienm}) or (\ref{peliniii}) by means of a $B$--transformation. Thus $k=0$ corresponds to a thermal description of phenomena, while $k=n$ corresponds to a quantum description of phenomena, no interpolation existing between the two descriptions. Nonextremal values of the type, {\it i.e.}\/, such that $0\neq k\neq n$, contain both thermal {\it and}\/ quantum descriptions simultaneously.

Average values $\langle f\rangle$ of functions $f$ on generalised complex manifolds are defined by an obvious modification of the product of the right--hand sides of Eqs. (\ref{septimo}) and (\ref{severus}).

\section{When ``quantum" becomes ``thermal"}\label{wqbt}

Any gravitational field is locally equivalent to an accelerated frame. In an accelerated frame, {\it quantum}\/ becomes {\it thermal}; this is basically the content of the Unruh effect \cite{UNRUH} (in an admittedly lax formulation that is however precise enough for our purposes). Without using the full apparatus of relativistic quantum field theory, let us see how {\it quantum}\/ can become {\it thermal}\/ in the simplified setup of the quantum mechanics of a nonrelativistic particle. This understood, we will analyse the role played by the GCS on phase space under the passage from an inertial frame to an accelerated frame.  We will conclude that the transformation law for the Schroedinger wavefunction under the passage to a noninertial frame (as in the Unruh effect) is governed by a $B$--transformation of the GCS on phase space. 

A remark is in order. The gravitational field considered here must be weak in order to rule out effects such as, {\it e.g.}\/,  relativistic speeds, or the likely breakdown of standard quantum mechanics in the presence of very strong gravitational fields \cite{PENROSE}. Such phenomena lie beyond our scope.

\subsection{Inclusion of a gravitational field}

In flat Euclidean space $\mathbb{R}^3$, let $K$ denote an inertial frame with origin $O$ and axes $Ox$, $Oy$ and $Oz$. 
Let $K'$ denote a uniformly accelerated frame, with origin $O'$ and axes $O'x'$, $O'y'$ and $O'z'$ respectively parallel to $Ox$, $Oy$ and $Oz$.
For simplicity we will assume that, at $t=0$, the two origins $O$ and $O'$ coincide, their relative velocity also vanishing at $t=0$. Let the acceleration $\vec{\alpha}$ of $K'$ with respect to $K$ be $(\alpha,0,0)$, with $\alpha$ a constant. Coordinates $(x,y,z)$ with respect to $K$ are related to coordinates $(x',y',z')$ with respect to $K'$ as per
\begin{equation}
x=x'+\frac{1}{2}\alpha t^2, \qquad y=y', \qquad z=z',\qquad t=t'.
\label{gala}
\end{equation}
We consider a point particle of mass $m$ fixed to the origin $O'$, thus at rest with respect to $K'$. If $H$ denotes the Hamiltonian of the particle as seen from the inertial frame $K$, then the Hamiltonian $H'$ in  $K'$ reads
\begin{equation}
H'=H-p_x\alpha t+\frac{m}{2}\alpha^2 t^2,
\label{laghache}
\end{equation}
with the momenta $p_x$ and $p'_x$ related as per $p_x'=p_x-m\alpha t$. In the inertial frame $K$ we have a Schroedinger equation ${\rm i}\hbar\partial \psi/\partial t=H\psi$. Our aim is to derive a transformation law for the wavefunction $\psi$ such that, in the accelerated frame $K'$, the Schroedinger equation will read ${\rm i}\hbar\partial \psi'/\partial t=H'\psi'$. For this purpose let us make the Ansatz
\begin{equation}
\psi'=\psi\exp\left[f(t)\right], 
\label{ansas}
\end{equation}
$f(t)$ being an undetermined function of the time variable. In this way we arrive at the following differential equation for the unknown function $f$: 
\begin{equation}
{\rm i}\hbar\frac{{\rm d}f}{{\rm d}t}=-p_x\alpha t+\frac{1}{2}m\alpha^2 t^2.
\label{diferenci}
\end{equation}
Dropping an irrelevant integration constant and substituting the result into Eq. (\ref{ansas}) leads to 
\begin{equation}
\psi'=\exp\left[-\frac{{\rm i}}{\hbar}\left(\frac{1}{6}m\alpha^2t^3-\frac{1}{2}p_x\alpha t^2\right)\right]\psi.
\label{mengen}
\end{equation}
Clasically, the particle is at rest in the frame $K'$, so $p'_x=0$ implies $p_x=m\alpha t$. Quantum--mechanically we can only state that the centre of mass remains at rest at $x'=0$, the wavepacket spreading around this average position. With this understanding we can also set $\langle p_x\rangle=p_x=m\alpha t$ in Eq. (\ref{mengen}). We conclude that, taking the wavefunction in the accelerated frame to be 
\begin{equation}
\psi'=\exp\left(\frac{{\rm i}}{\hbar}\frac{1}{3}m\alpha^2t^3\right)\psi
\label{insieme}
\end{equation}
ensures the form invariance of the Schroedinger equation under the transformation from an inertial frame to an accelerated frame. For time lapses that are short enough, and/or for accelerations that are weak enough, the speeds attained will never become relativistic. Within this limited range, Newtonian mechanics (and its quantum counterpart, the Schroedinger equation) can be trusted.

\subsection{The Unruh effect}

The next step is to invoke de Broglie \cite{DEBROGLIE} in order to write an inverse proportionality between time $t$ and temperature $T$:
\begin{equation}
-\frac{{\rm i}}{t}=\frac{k_B}{\hbar}T.
\label{tiemponuevo}
\end{equation}
Thus substituting Eq. (\ref{tiemponuevo}) into (\ref{insieme})  we find
\begin{equation}
\psi'=\exp\left(-\frac{1}{3}\frac{m\alpha^2\hbar^2}{k_B^3T^3}\right)\psi.
\label{fiducia}
\end{equation}
Moreover, from the above we can read off what power law must relate the acceleration to the temperature of the accelerated frame: $\alpha$ must be proportional to $T$, while dimensional analysis provides the necessary conversion factors. Specifically,
\begin{equation}
\alpha=2\pi\frac{ck_B}{\hbar}T.
\label{termo}
\end{equation}
The dimensionless normalisation factor $2\pi$, that cannot be derived using our simplified treatment, comes from a full quantum--field--theoretical analysis \cite{UNRUH}. Finally substituting Eq.  (\ref{termo}) into (\ref{fiducia}) we arrive at 
\begin{equation}
\psi'=\exp\left(-\frac{4\pi^2}{3}\frac{mc^2}{k_BT}\right)\psi.
\label{regius}
\end{equation}
Eqs. (\ref{regius}) and (\ref{insieme}) are equivalent, the equivalence between the two being guaranteed by the de Broglie relation (\ref{tiemponuevo}) and the Unruh relation (\ref{termo}). 

The Boltzmann--like factor present in Eq. (\ref{regius}) bears out the fact that the effect of the gravitational field on the Schroedinger wavefunction is of thermal nature. Due to the assumptions made in our derivation, Eq. (\ref{regius}) is valid only for intermediate temperatures. The limit $T\to\infty$  is excluded (because this would require strong gravitational fields); so is the limit $T\to 0$ (because of the inverse proportionality (\ref{tiemponuevo}) between time and temperature).

\subsection{Transformation to an accelerated frame as a $B$--transformation}

Classical phase space is spanned by the coordinates $x,y,z$ and their conjugate momenta $p_x,p_y,p_z$. For the rest of the discussion, the dimensions $y$, $p_y$, $z$, $p_z$ can be ignored, as they are unaffected by the change of frame (\ref{gala}). Thus, for our purposes, the manifold $\mathbb{M}$ of section \ref{jichin} can be taken to be that subspace of classical phase space spanned by $x$ and $p_x$, {\it i.e.}\/, $\mathbb{R}^2$.

Now the manifold $\mathbb{R}^2$ can be endowed with a GCS. This can be done in two equivalent ways. One can consider the GCS of complex type defined on $\mathbb{R}^2=\mathbb{C}$ by the complex coordinates (\ref{koomfgerl}). Alternatively, one can consider the GCS of symplectic type defined on $\mathbb{R}^2$ by the standard symplectic form $\omega={\rm d}x\wedge{\rm d}p_x/\hbar$. Since our interest lies in considering the effect of $B$--transformations, and $\mathbb{R}^2=\mathbb{C}$ is a K\"ahler manifold, the type of the CGS considered is immaterial.

We claim that the transformation law for the Schroedinger wavefunction under the passage to an accelerated frame, Eq. (\ref{insieme}), follows from a $B$--transformation of the GCS on phase space $\mathbb{R}^2$, Eq. (\ref{azaazo}). In other words, the Schroedinger wavefunction keeps track of which frame is being used, the bookkeeping device being the GCS on phase space. Verifying that such is indeed the case requires, so to speak, translating the geometer's language into the physicist's language. This we do next.

Tangent vectors $X$ at the point $(x,p_x)\in\mathbb{R}^2$ are objects 
\begin{equation}
X=a\partial_x+ b\partial_{p_x}\in T_{(x,p_x)}\mathbb{R}^2,\qquad a,b\in\mathbb{R}.
\label{xam}
\end{equation}
Similarly, tangent covectors $\xi$ at the point $(x,p_x)\in\mathbb{R}^2$ are objects 
\begin{equation}
\xi=c{\rm d}x+d{\rm d}p_x\in T^*_{(x,p_x)}\mathbb{R}^2,\qquad c,d\in\mathbb{R}.
\label{marj}
\end{equation}
As the basepoint $(x,p_x)\in\mathbb{R}^2$ is moved around, we obtain a vector field $X$ and a field of differential 1--forms $\xi$ on $\mathbb{R}^2$.
This amounts to promoting the numbers $a,b,c,d$ to real--valued functions $a(x,p_x)$, $b(x,p_x)$, $c(x,p_x)$, $d(x,p_x)$ on $\mathbb{R}^2$. 
Finally, an object such as $X+\xi$ in Eq. (\ref{azaazo}) is the direct sum of a vector field and a field of differential 1--forms on $\mathbb{R}^2$---a section of the direct sum bundle $T\mathbb{R}^2\oplus T^*\mathbb{R}^2$.

Next we reexpress the $B$--transformation (\ref{azaazo}) as the variation
\begin{equation}
\delta (X+\xi)=\delta X+\delta \xi = \delta\xi=i_XB.
\label{trafog}
\end{equation}
Above we have used the fact that, under a $B$--transformation, $X$ remains unchanged. The $B$--field is a 2--form on $\mathbb{R}^2$, 
\begin{equation}
B=B(x,p_x){\rm d}x\wedge{\rm d}p_x,
\label{ipsa}
\end{equation}
with a certain coefficient function $B(x,p_x)$. Now 
\begin{equation}
\delta\xi=i_XB=a(x,p_x)B(x,p_x){\rm d}p_x+b(x,p_x)B(x,p_x){\rm d}x.
\label{kontrr}
\end{equation}
The above is a 1--form field, so it can be added to $X+\xi$ as required by Eq. (\ref{azaazo}). Let us now make the following specific choice for the vector field $X$:
\begin{equation}
a(x,p_x)=x, \qquad b(x,p_x)=p_x.
\label{pottss}
\end{equation}
In the physicist's language, this $X$ is just the position vector on phase space $\mathbb{R}^2$. Substituted into Eq. (\ref{kontrr}), this choice for $X$ yields
\begin{equation}
\delta\xi=i_XB=xB(x,p_x){\rm d}p_x+p_xB(x,p_x){\rm d}x.
\label{yyeldman}
\end{equation}

Along the motion of the particle located at $O'$ we can write, using Eq. (\ref{gala}),
\begin{equation}
{\rm d}p_x=m\alpha{\rm d}t, \qquad {\rm d}x=\alpha t{\rm d}t.
\label{solascriptura}
\end{equation}
Substitution of Eqs. (\ref{gala}) and (\ref{solascriptura}) into (\ref{yyeldman}) leads to
\begin{equation}
\delta\xi=i_XB=\frac{3}{2}B(x(t),p_x(t))m\alpha^2 t^2{\rm d}t.
\label{solagratia}
\end{equation}
The above is a 1--form, that can be integrated along the trajectory followed by the particle between $\tau=0$ and $\tau=t$. We denote by $\Delta\xi(t)$ the number so obtained:
\begin{equation}
\Delta\xi(t):=\int_0^t\delta \xi=\frac{3}{2}m\alpha^2\int_0^tB(x(\tau),p_x(\tau)) \tau^2{\rm d}\tau.
\label{tresolas}
\end{equation}
When $B$ is a constant, the integral can be evaluated explicitly:
\begin{equation}
\Delta\xi(t)=\frac{1}{2}Bm\alpha^2 t^3.
\label{bariazia}
\end{equation}
That the function $B(x(t),p_x(t))$ is actually constant on $\mathbb{R}^2$ implies that the 2--form $B$ in Eq. (\ref{ipsa}) becomes a mere scalar multiple of the canonical symplectic form on phase space. Specifically, picking $B=2/3$ we find in (\ref{bariazia})
\begin{equation}
\Delta\xi(t)=\frac{1}{3}m\alpha^2 t^3.
\label{geizistgeil}
\end{equation}
The right--hand side of the above equals ($-{\rm i}\hbar$ times) the argument of the exponential in the Unruh transformation law (\ref{insieme}). Therefore the latter can be reexpressed as
\begin{equation}
\psi'=\exp\left(\frac{\rm i}{\hbar}{\Delta\xi}(t)\right)\psi.
\label{loewe}
\end{equation}

Summarising, we may say that {\it the Unruh effect acts on the wavefunction by multiplication with the exponential of (${\rm i}/\hbar$ times) the integral of a $B$--field along the particle's trajectory on phase space}\/. The vector field $X$ involved in this $B$--transformation is just the position vector on phase space, while the $B$--field considered is a mere scalar multiple of the canonical symplectic form on phase space.

\subsection{A  nonuniform gravitational field}

The relation just derived between the Unruh effect and the $B$--transformation of the GCS on phase space was based on the assumption that the gravitational field was static and spatially constant. In turn, this assumption made it possible to choose a constant $B$--field on phase space (actually a scalar multiple of the symplectic form). A nonstatic and/or nonuniform gravitational field can be mimicked by a nonstatic and/or nonuniform acceleration vector $\vec{\alpha}$. This lends plausibility to the following hypothesis:

{\it Regard classical phase space as a generalised complex manifold. In the presence of a nonstatic and/or nonuniform, but nevertheless weak, gravitational field, the inertial--frame Schroedinger wavefunction $\psi$ remains form--invariant under a transformation to a locally accelerated frame, where its value is $\psi'$, provided that $\psi$ and $\psi'$ are related according to the law
\begin{equation}
\psi'=\exp\left(\frac{\rm i}{\hbar}\Delta\xi(t)\right)\psi.
\label{papalapapp}
\end{equation}
Above, 
\begin{equation}
\Delta\xi(t):=\int_0^ti_XB(x(\tau),p_x(\tau)){\rm d}\tau
\label{winkle}
\end{equation}
is a line integral along the particle's trajectory in phase space, while $X$ is the position vector of the particle along the said trajectory. Moreover, whenever the generalised complex structure on classical phase phase is of symplectic type, the 2--form $B$ is an appropriate scalar multiple of the  symplectic form $\omega$}\/.

We defer analysis of the above hypothesis for further study.

\section{Conclusions}\label{folinlab}

We have presented a brief review of some recent developments in differential geometry with applications to thermodynamical fluctuation theory.  Standard wisdom draws a clear frontier between {\it thermal}\/ fluctuations and {\it quantum}\/ fluctuations. While this separation is perfectly consistent in the absence of gravitational fields, this border becomes fuzzy in the presence of gravity \cite{PADDY1, PENROSE, SMOLIN1, SMOLIN2}. A well--known example of this mixing is the Unruh effect \cite{DAVIES, FULLING, UNRUH}. Another instance of a gravitational incursion into the thermal realm is the Ehrenfest--Tolman effect \cite{TOLMAN}. One can expect an eventual theory of quantum gravity to enhance, rather than diminish, this mixing of thermal and quantum phenomena.

In this article we have examined the thermalising effect of  weak, classical gravitational fields on the Schroedinger wavefunction from the point of view of generalised complex geometry on classical phase space. Using the transformation law for the Schroedinger wavefunction under the passage to an accelerated frame, we have derived the nonrelativistic Unruh effect. As expected, the latter establishes a linear dependence law between the acceleration of the noninertial frame and the temperature thereby generated. Within the scope of the techniques presented here lie other interesting physical systems, to be treated in an upcoming publication. Such are quantum--classical hybrids \cite{ELZE1, ELZE2} and the thermalising properties of nonuniform (but still weak and classical) gravitational fields.

Altogether, we conclude that generalised complex geometry provides a powerful tool to analyse fluctuation theory and thermal phenomena in the presence of gravity.  
\vskip0.5cm
\noindent
{\it  Nihil sapientiae odiosius acumine nimio}\/---Seneca.

\end{document}